\documentclass[aps,prl,twocolumn,showpacs,superscriptaddress]{revtex4}
\usepackage{amsmath}
\usepackage{amssymb}
\usepackage{epsfig}
\usepackage{color}
\usepackage{graphicx}
\usepackage{graphicx,amsmath}
\usepackage{bm}
\usepackage{epstopdf}

\begin{document}
\newcommand{\beq}{\begin{equation}}
\newcommand{\eeq}{\end{equation}}

\title{High-$T_c$ superconductivity of electron systems with flat bands
pinned to the Fermi surface}
\author{V.~A. Khodel}
\affiliation{NRC Kurchatov Institute, Moscow, 123182, Russia}
\affiliation{McDonnell Center for the Space Sciences \& Department of Physics,
Washington University, St.~Louis, MO 63130, USA}
\author{J.~W. Clark}
\affiliation{McDonnell Center for the Space Sciences \& Department of Physics,
Washington University, St.~Louis, MO 63130, USA}
\affiliation{Centro de Ci\^encias Matem\'aticas, Universidade de Madeira,
9000-390 Funchal, Madeira, Portugal}
\author{V.~R. Shaginyan}
\affiliation{Petersburg Nuclear Physics Institute, NRC Kurchatov Institute,
Gatchina, 188300, Russia}
\affiliation{Clark Atlanta University, Atlanta, GA 30314, USA}
\author{M.~V. Zverev}
\affiliation{NRC Kurchatov Institute, Moscow, 123182, Russia}
\affiliation{Moscow Institute of Physics and Technology, Dolgoprudny, Moscow
District 141700}

\begin{abstract}

The phenomenon of flat bands pinned to the Fermi surface is analyzed on
the basis of the Landau-Pitaevskii relation, which is applicable to
electron systems of solids.  It is shown that the gross properties
of normal states of high-$T_c$ superconductors, frequently called strange
metals, are adequately explained within the flat-band scenario.  Most
notably, we demonstrate that in electron systems moving in a two-dimensional
Brillouin zone, superconductivity may exist in domains of the Lifshitz
phase diagram lying far from lines of critical antiferromagnetic fluctuations,  even if the effective electron-electron interaction in the Cooper channel
is repulsive.

\end{abstract}

 \maketitle

A central issue confronting present-day condensed matter theory is that
of unambiguous determination of the mechanisms governing the rich
non-Fermi-liquid (NFL) behavior revealed by intensive experimental studies
of strongly correlated electron systems of solids and liquid $^3$He films.
Prominent on the scene of this seminal area of condensed matter physics
are numerous versions of the Hertz-Millis-Moriya (HMM) description.
These scenarios ascribe such NFL behavior to quantum critical
fluctuations.  Proponents of the HMM approach claim that accounting for
such fluctuations also allows one to explain the phenomenon of high-$T_c$
superconductivity, which entails substantial enhancement of the
critical temperatures $T_c$ of superconducting phase transitions
relative to standard BCS values.

In BCS theory, a typical $T-x$ phase diagram of a
superconducting system ($x$ being a relevant control
parameter such as doping or pressure) consists of an
''island'' of superconductivity, surrounded by a
Fermi-liquid (FL) ''sea,'' a domain where the resistivity
varies as $\rho_{FL}(T)=\rho_0+A_2T^2$.  Real phase
diagrams of strongly correlated metals exhibiting
superconductivity look different, as exemplified by the
phase diagram of the LCCO family of electron-doped
high-$T_c$ superconductors. This phase diagram, established
empirically in Refs.~\cite{greene,butch}, is reproduced in
Fig.~1. A prominent feature is the presence of two distinct
regimes of NFL temperature behavior of the resistivity
$\rho(T)$ at $T>T_c$, which separate the ''blue FL sea''
from the ''yellow island'' of superconductivity.  In the
interval $T_c(x)<T<T_1(x)$, the resistivity $\rho(T)$
changes linearly with $T$, thus $\rho(T)=\rho_0+A_1T$.
Above $T_1(x)$ a different NFL regime appears, in which
$\rho(T)$ varies as $T^n$ with $n\simeq 1.6$.

In the conventional HMM approach, the NFL linearity of the
resistivity $\rho(T)$ observed in normal states of many
high-$T_c$ compounds is attributed to antiferromagnetic
critical fluctuations with wave vector ${\bf
Q}=(\pi/a,\pi/a)$. In two-dimensional (2D) systems of
electrons moving in the external field of a square lattice,
this explanation works provided saddle points are located
close to the Fermi lines; otherwise it {\it fails}.
Significantly, the additional NFL regime in the LCCO phase
diagram of Fig.~1 having $n \simeq 1.6$ in
$\rho(T)=\rho_0+A_n T^n$ terminates at the same critical
doping as the linear-in-$T$ regime.  This behavior cannot
plausibly be associated with antiferromagnetic fluctuations
\cite{armitage}. Further, there are numerous examples of 3D
systems whose low-$T$ resistivity also varies linearly with
$T$, in contrast to predictions of the spin-fluctuation
scenario for 3D systems. In addition, fluctuation-induced
phenomena such as critical opalescence, exhibited as a huge
enhancement in absorption of light by a liquid that is
ordinarily transparent, emerge only in the {\it immediate
vicinity} of points (or lines) of second-order phase
transitions. The range $\Delta T$ of the interval $T_N-T$
impacted by critical fluctuations on the ordered side of
the transition is determined by equating the mean-field
value of the order parameter, behaving as $\sqrt{T_N-T}$,
to the corresponding fluctuation contribution.
Consequently, the fluctuation scenarios become irrelevant
when $|T-T_N| > \Delta T$.

In fact, experiments on the systems of interest often furnish
clear evidence for the persistence of NFL behavior {\it far}
from such lines of criticality.  In the well-studied heavy-fermion
metal YbRh$_2$Si$_2$ \cite{steglich}, which provides the classic
example of NFL linear behavior of $\rho(T)$ in the disordered phase
{\it and} a $T^2$ resistivity regime on the ordered side of the
posited antiferromagnetic phase transition, the latter behavior
is found to prevail {\it almost up to the transition temperature}
$T_N=70$ mK.  Given that the FL regime of $\rho(T)$ holds in the
antiferromagnetic state of YbRh$_2$Si$_2$ almost up to $ T_N$,
the critical fluctuations must be weak; otherwise the linear-in-$T$
corrections to $\rho_{FL}(T) $ due to these fluctuations would
be significant on the ordered side as well.  However, their weakness
stands in blatant contradiction to the linear-in-$T$ variation of the
resistivity on the disordered side, which is maintained from
$T_N=70\, mK $ up to $T>1~K$ \cite{steglich}.  The inescapable
conclusion is that the proposed fluctuation mechanism is irrelevant
to the NFL behavior of $\rho(T>T_N)$ in this compound.

This example is not alone.  There are multiple instances of other
materials in which the FL behavior of $\rho(T)$ persists solely on
the ordered side of the transition, whereas on the disordered side the
resistivity changes linearly with $T$ -- despite the expectation that
FL theory should be {\it more} secure there than in the ordered state.
To be specific, consider the $T-P$ phase diagram of the strongly
correlated heavy-fermion metal CeCoIn$_5$\cite{sidorov}, presented in
Fig.~2.  A remarkable feature seen in this diagram is the crossover
from the NFL linear-in-$T$ regime of $\rho(T)$ to its FL regime,
occurring in the normal state far from the transition point.  This
process is accompanied by a dramatic change of the residual resistivity
$\rho_0$, which drops by factor around ten on the FL side of the
crossover \cite{sidorov}.  Such behavior is inconceivable within
the textbook understanding of kinetic phenomena in Fermi liquids.

All these facts and many others portend that in some strongly correlated
electron systems of solids, it is the single-particle degrees of
freedom that are the real playmakers in the observed NFL behavior,
rather than critical fluctuations.
By implication the HHM approach is fallacious when applied
to these systems, since the single-particle degrees are
integrated out.

It has become clear that the distinctive signature of the underlying
physics of the Lifshitz phase diagram is the appearance of a so-called
quantum critical point (QCP) where the density of states $N(T=0)$,
associated in homogeneous matter with the effective mass $M^*$, is
{\it divergent}. Accordingly, the Landau state becomes unstable
beyond the QCP and necessarily undergoes rearrangement \cite{physrep}.
The relevance of the QCP to high-$T_c$ superconductivity has been
confirmed in recent experimental work of Ramshaw et al. \cite{Science2015}
on YBa$_2$Cu$_3$O$_{6+x}$, one of the most prominent high-$T_c$
superconductors, with $T_c$ around $90\, {\rm K}$. In documenting
''A quantum critical point at the heart of high-$T_c$ superconductivity,''
they have studied the enhancement of the electron effective mass $M^*$
in dHvA oscillations of the normal metallic state, subjecting the
sample to huge magnetic fields $B$ of more than $90\, {\rm K}$ that
terminate superconductivity.  The measured effective mass $M^*(B)$ is
found to {\it diverge} at the optimal doping $x_o$, where the critical
temperature $T_c(x,B=0)$ attains its maximum.

It is noteworthy that the presence of a QCP in the Lifshitz phase
diagram of homogeneous 3D electron systems was predicted in
microscopic parameter-free calculations \cite{zks} prior to
its experimental discovery in strongly correlated heavy-fermion
metals \cite{steglich}.  Application of the approach of Ref.~\cite{zks}
to the 2D problem \cite{bz} demonstrated that in this case the QCP
lies at realistic electron densities, corresponding to $r_s\simeq 8$.

Absent any change of symmetry, the anticipated rearrangement of the
Landau state occurring at the QCP is naturally attributed to
some topological transition.  The earliest topological scenario for
NFL behavior in strongly correlated Fermi systems, advanced more
than 20 years ago \cite{ks,noz,vol}, traced this behavior  to
an {\it interaction-induced} rearrangement of the Landau state,
often called fermion condensation (FC).  This phenomenon, described
more vividly as a swelling of the Fermi surface, is associated
with the occurrence of a flat band pinned to the Fermi surface.

In a significant formal development, the FC phenomenon was rediscovered
in 2009 within the framework of the adS-CFT duality \cite{lee}.  More
to the point, the formation of flat bands has been demonstrated both
analytically and numerically for the Hubbard model, one of the most
popular models of strongly correlated electron systems \cite{kats2014}.
Therefore the question of principle whether the FC rearrangement
exists and is relevant to condensed matter theory already has a
positive answer.  The remaining issue, addressed herein and elsewhere,
is whether or not the FC phenomenon can actually provide the basis
for a satisfactory explanation of the experimentally observed NFL behavior
of such systems.  That the FC scenario competes favorably with other
attempts to explain the salient experimental data has been established
in many studies, notably Refs.~\cite{shagrep, mig100,an2012,book}.
Additionally, invocation of FC theory has recently resolved a
long-standing puzzle associated with the disappearance of a specific
set of Shubnikov-de Haas magnetic oscillations in the 2D electron gas
of MOSFETs, which results in the doubling of oscillation periods
near a quantum critical point \cite{krav2000,shashkin}.

The paramount objective of this communication is to apply the flat-band
scenario to the elucidation of high-$T_c$ superconductivity, discovered
30 years ago and still a challenge to theoretical understanding.
The adequacy of any theory of this phenomenon depends largely on how
well it reproduces the properties of strange metals -- normal states
of high-$T_c$ superconductors.  The Landau FL theory of normal states
provides the basis for BCS theory, which, however, fails to describe
high-$T_c$ superconductivity.  With this in mind, we present the essential
elements of FC theory, focusing on those departures from FL theory
that may qualify it as a basis for understanding the phase diagrams
of high-$T_c$ materials.

It should be emphasized that FL and FC theories are both rooted in
seminal work of Landau \cite{lan} in that they employ the quasiparticle
formalism.  Hence they are applicable provided the damping of single-particle
excitations is small compared with their energy. In the Landau theory
of conventional Fermi liquids, this requirement is always met toward $T=0$,
since the damping is proportional to $T^2$.  The situation is more
complicated in the flat-band scenario, because systems having flat bands
belong in fact to the class of marginal Fermi liquids, in which the
damping changes linearly with $T$, but with a prefactor proportional
to the ratio $\eta$ of the volume occupied in momentum space by the
flat bands to the total Fermi volume (see below).  Thus, the above
requirement for applicability is met provided $\eta$ is small.

Importantly, in electron systems of solids where translational invariance
breaks down, the single-particle states are identified by quasimomentum
${\bf p}$, and the FL relation
\beq
n({\bf p})=(1+e^{\epsilon( {\bf p})/T})^{-1},
\label{nsp}
\eeq
between the quasiparticle momentum distribution $n({\bf p})$ and the
single-particle spectrum $\epsilon({\bf  p})$ (measured from the chemical
potential $\mu$), continues to apply.  As shown in Ref.~\cite{Mig90},
the Landau-Pitaevskii (LP) identity can be employed as a second relation
between these quantities that holds for the electron system moving
in the external field of the crystal lattice.  In the notation adopted
here, the LP equation takes the form
\beq
{\partial \epsilon({\bf p})\over \partial {\bf p}}={\partial\epsilon_0({\bf p})
\over\partial {\bf p}}+\int f({\bf p},{\bf p}')
{\partial n({\bf p}')\over \partial {\bf p}'}d\upsilon'  ,
\label{lp}
\eeq
where the quantity $ \partial \epsilon_0({\bf p})/\partial {\bf p}$
contains only regular contributions coming from domains far from the
Fermi surface.

Like the corresponding set of equations for homogeneous
matter the set (\ref{nsp}) and (\ref{lp}) also possesses a
class of flat-band solutions for which the electron group
velocity vanishes in a finite domain ${\bf p}\in \Omega$.
Accordingly, one is required to solve the reduced equation
\beq 0={\partial\epsilon_0({\bf p})\over\partial {\bf p}}
+\int f({\bf p},{\bf p}') {\partial n_*({\bf p}') \over
\partial {\bf p}'}d\upsilon', \quad {\bf p},{\bf p}'\in \Omega .
\label{grfc}
\eeq
in this region.  The function $n_*({\bf p})$ is introduced to represent
a nontrivial FC solution of this equation.  Outside the FC region,
the quasiparticle spectrum obeys the familiar Landau-type relation
\beq
{\partial \epsilon({\bf p})\over \partial {\bf p}}=
{\partial\epsilon_0({\bf p})\over\partial {\bf p}}
+\int f({\bf p},{\bf p}') {\partial n_*({\bf p}')\over
\partial {\bf p}'}d\upsilon' ,\quad {\bf p} \notin \Omega .
\label{grb}
\eeq
Such a solution is characterized by its topological charge (TC),
an invariant expressed in terms of a contour integral
constructed from the single-particle Green function $G({\bf p},\varepsilon)$
and its derivatives \cite{vol,volovik}.  The TC of a state exhibiting
a flat band takes a {\it half-odd-integral} value, whereas the TC
assigned to a Lifshitz state featuring a multi-connected Fermi
surface (or ``Lifshitz pockets'') is always integral, since this
state has standard quasiparticle occupation numbers $n({\bf p})=0,1$.

Proceeding to low $T\neq 0$, one may insert the solution $n_*({\bf p})$
of Eq.~(\ref{grfc}) as a zeroth approximation for $n({\bf p},T)$ to
obtain \cite{noz}
\beq
\epsilon({\bf p},T\to 0)=T\ln {1-n_*({\bf p})\over n_*({\bf p})} ,\quad
{\bf p}\in \Omega .
\label{noze}
\eeq
In the FC momentum region ${\bf p}\in \Omega $, the resulting dispersion
$v_n=\partial \epsilon({\bf p})/\partial p_n$ of the single-particle
spectrum is then found to be proportional to $T$.

An essential feature of the flat-band scenario is the presence of a
residual entropy \cite{ks,an2012}
\beq
S_*=-\!\!\int\limits_\Omega [(1{-}n_*({\bf p}))\ln(1{-}n_*({\bf p})) +n_*({\bf p})\ln n_*({\bf p})]d\upsilon
\eeq
associated with the FC region, where the occupation numbers $n_*({\bf p})$
differ from 0 and 1.  This residual entropy does not contribute
at all to the specific heat $C(T)=TdS/dT$.  However, in normal states
of systems with flat bands, $S_*(T\to 0)$ retains a finite value
and makes a huge $T$-independent contribution to the thermal expansion
$\beta\propto \partial S/\partial P$ of these states. This conclusion
is in agreement with the thermal expansion of the strongly correlated
heavy-fermion metal CeCoIn$_5$ measured at $T>T_c\simeq 2.3\ K$
\cite{oeschler}.

To avoid contradiction with the Nernst theorem mandating $S(0)=0$, the
residual entropy $S_*$ must be released by means of some first- or
second-order phase transitions, or with the aid of crossovers to a state
having a multi-connected Fermi surface formed by a set of Lifshitz pockets
with $n({\bf p})=0,1$, hence implying $S_*=0$ \cite{ks,an2012,book}. This
ramification explains the diversity of quantum phase transitions
observed in strongly correlated electron systems, stemming from the
interplay between antiferromagnetism, charge order, and superconductivity.

\begin{figure}[!ht]
\begin{center}
\includegraphics [width=0.47\textwidth]{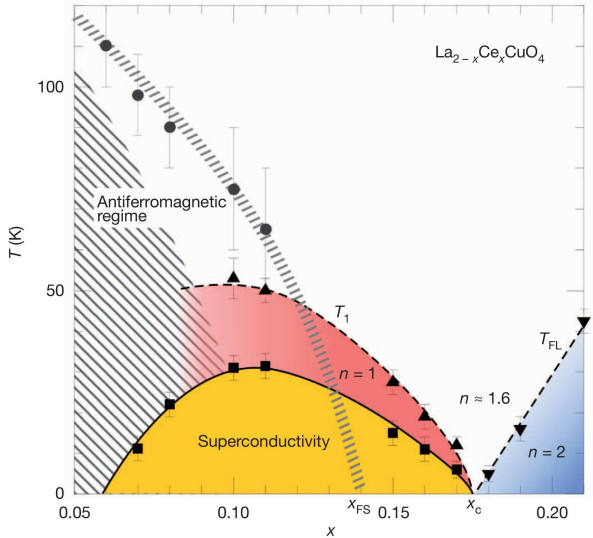}
\end{center}
\caption{(color online) Temperature-doping $T-x$ phase
diagram of La$_{2-x}$Ce$_x$CuO$_4$ \cite{greene}, reprinted
with authors' permission. The yellow region is the
superconducting dome. The resistivity $\rho(T)$ in the
different normal phases has the form
$\rho(T)=\rho_0+A_nT^n$, with $n=2$ for the FL domain
(blue), $n=1$ for the NFL FC domain (red), and $n=1.6$ for
the NFL QCP domain (white) ( see the text). The
temperatures $T_1$ (triangles) and $T_{FL}$ (inverted
triangles), traced by dashed curves, indicate the crossover
temperatures to the linear-in-$T$ regime of $\rho(T)$ and
the FL regimes, respectively.} \label{fig1}
\end{figure}

\begin{figure}[!ht]
\begin{center}
\includegraphics [width=0.47\textwidth]{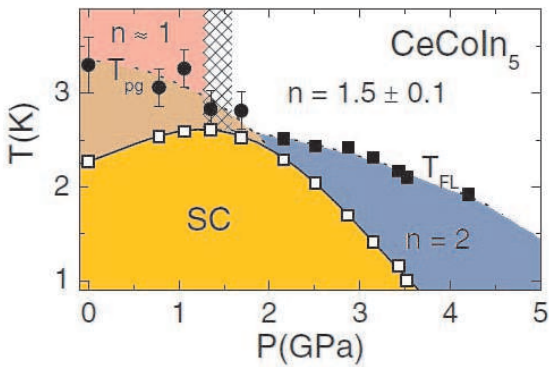}
\end{center}
\caption{(color online) Temperature-pressure $T-P$ phase diagram of
CeCoIn$_5$ \cite{sidorov}, constructed based on the phase diagram
published in Ref.~\cite{sidorov}.  Unlike the phase diagram of
Fig.~\ref{fig1}, there exists a small pseudogap (PG) region in
orange where superconductivity is precluded, but a gap $\Delta$
persists. The occurrence of a PG region in this compound will
be discussed in detail in a future article.}
\label{fig2}
\end{figure}

Let us examine more closely the NFL behavior of the resistivity
$\rho(T)$ at $T>T_c$ in superconducting materials exhibiting flat
bands, recognizing that this is the foundation of the phase diagrams
presented in Figs.~\ref{fig1} and \ref{fig2}.  Ideally, the
collision term differs from zero only due to Umklapp processes.
However, strongly correlated electron systems of solids have open
Fermi surfaces where these processes work in full force, such that
the detailed structure of the kernel can be ignored, and we are
left with the integral
\begin{eqnarray}
&I(n)\propto \int [n_1n_2(1{-}n_1')(1{-}n_2')-n_1'n_2'(1{-}n_1)(1{-}n_2)]
\quad \nonumber \\
&\times\delta ({\bf p}_1{+}{\bf p}_2{-}{\bf p}'_1{-}{\bf p}'_2)\,
\delta(\epsilon_1{+}\epsilon_2{-}\epsilon'_1{-}\epsilon'_2)\,d\upsilon_1
d\upsilon'_1d\upsilon_2d\upsilon'_2.
\label{coll}
\end{eqnarray}
Eq.~(\ref{coll}) exhibits several factors $1/v_n({\bf
p},T)$ upon making the standard replacement $d^3p\to dS
(d\epsilon/v_n({\bf p}))$, where $dS$ is an element of the
isoenergetic surface.  In conventional Fermi liquids, these
factors are $T$-independent and yield the FL result.
Contrariwise, in systems with flat bands, it is seen from
Eq.~(\ref{noze}) that the behavior $v_n({\bf p},T)\propto
T$ applies in the ``hot spot" associated with the
corresponding FC region.  Any momentum integration over
this region then contributes a factor $\eta/T$ to the
collision integral (and to the resistivity as well), where
$\eta$ is the dimensionless FC density.  This causes a
damping of single-particle excitations, rendering systems
with flat bands {\it marginal Fermi liquids} \cite{kss,kz}.
Such a conclusion is in agreement with the results of
experimental studies of kinetic properties of the
heavy-fermion metal CeCoIn5 \cite{sidorov,thompson2012}.

Ordinarily $\eta $ is very small, so we need only retain
the leading terms of order $\eta $ and $ \eta^2$ to arrive
at the resistivity expression \beq
\rho(T)=\rho_0(P,x)+A_1(P,x) T \label{nfl} \eeq in the
presence of a flat band, with \beq
\rho_0(P,x)=\rho_0^i+a_0\eta^2(P,x), \quad A_1(P,x)=a_1
\eta(P,x), \label{a1} \eeq wherein $\rho_0^i$ denotes an
impurity-induced contribution to the residual resistivity
$\rho_0$ and $a_0$, $a_1$ are constants.  The total
residual resistivity $\rho_0$ becomes dependent on pressure
$P$ and doping $x$.  The cleaner the metal, the greater the
magnitude of the jump of the residual resistivity $\rho_0$
in the rapid crossover from the NFL regime (\ref{a1}) to
the standard FL regime. In the limit $\rho_0^i\to 0$, the
ratio of the $\rho_0$ value on the NFL side of the
crossover to that on the FL side tends {\it to infinity}.
This observation serves to explain the rapid variation of
$\rho_0$ found in the especially clean samples of
CeCoIn$_5$ \cite{sidorov}.  Such otherwise puzzling
behavior has also been reported in measurements of the
residual resistivity $\rho_0(P)$ of the metal CeAgSb$_2$,
where the aforesaid ratio attains huge values of order
$10^2$ \cite{onuki}.

As will be demonstrated below (cf.\ Eq.~(\ref{reqh})), the critical
temperature $T_c$ for termination of superconductivity in high-$T_c$
superconductors is proportional to the FC density $\eta$, so that
\beq
A_1(x)\propto T_c(x)\propto \eta(x).
\label{prop}
\eeq
Hence the theoretical ratio $T_c/A_1$ is independent of the FC density.
While $\eta$
changes somehow with doping $x$, the ratio $T_c/A_1$ turns out
to be independent of $x$, in agreement with the experimental
behavior uncovered in the electron-doped materials LCCO and PCCO,
as well as in the Bechgaard class of organic superconductors
(TMTSF)$_2$PF$_6$ \cite{greene,taillefer}.  In addition, the
relation (\ref{prop}) implies that the factor $A_1$, which specifies
the linear-in-$T$ NFL regime of resistivity, vanishes at the
same doping $x_c$ where high-$T_c$ superconductivity terminates,
again in agreement with experiment (see Fig.~\ref{fig1}).  Thus we
infer that three different resistivity regimes come into play in the
immediate vicinity of FC onset: (i) the FL regime $\rho(T) \propto T^2$,
(ii) the FC regime $\rho(T) \propto T$, and (iii) the high-$T_c$
superconducting regime with $\rho=0$.

On the other hand, for heavy-fermion superconductors such as CeCoIn$_5$
in which the critical temperature $T_c$ is extremely low, the
BCS logarithmic term cannot be ignored.  In this situation, the onset of
FC is disconnected from the occurrence of superconductivity, and
merging of different resistivity regimes in the corresponding normal
states does not take place. This conclusion is seen to be in agreement
with the $T-P$ phase diagram of CeCoIn$_5$ displayed in Fig.~\ref{fig2}.

In the systems under study, still another profound NFL contribution
to $\rho(T) $ is possible, specific to what can be called the QCP
resistivity regime. Elucidation of its emergence is especially simple
in the homogeneous electron liquid where, at the QCP, the Fermi
velocity vanishes to yield
\beq
\epsilon(p\to p_F)\propto (p-p_F)|p-p_F|
\label{qcpe}
\eeq
and hence $v(p)=d\epsilon(p)/dp\propto \sqrt{|\epsilon(p)|}$.
One easily verifies that the leading contribution to $\rho(T)$
now increases as $T^{3/2}$.  This distinctive NFL regime has its
onset {\it at the  critical doping} $x_c$ as well.  It exists in
those electron systems that possess a QCP, e.g.\ in the LCCO
family \cite{greene} and in CeCoIn$_5$ \cite{paglione}.  There a
minor difference between our theoretical predictions and experiment.
In the white NFL regime of Fig.~\ref{fig1}, the measured resistivity
$\rho(T)$ varies as $T^{1.6}$, and in the corresponding regime
of Fig.~\ref{fig2}, as $\rho(T)\propto T^{1.5\pm 0.1}$.  Our
analysis leads to the relation $\rho(T)=A_{3/2}T^{3/2} +A_2T^2$.

Having confirmed that the flat-band-scenario can successfully explain
gross properties of the strange metals, we may now turn to the
primary aim of this article: explanation of the occurrence of
$D$-pairing and dramatic enhancement of its critical temperature $T_c$.
In doing so we will proceed within the standard BCS theory, ignoring
an enigmatic pseudogap phenomenon.  In this case, the structure of
the gap function and the magnitude of $T_c$ are revealed with the
aid of the linearized BCS gap equation
\beq
\Delta({\bf p}) =  - \int {\cal V}({\bf p},{\bf p}')
\tanh \left({\epsilon({\bf p}',T_c)\over  2T_c}\right){\Delta ({\bf p}')
\over 2|\epsilon({\bf  p}',T_c)| }d\upsilon' .
\label{bcst}
\eeq
Upon defining the functions
$$
X({\bf p},T_c) = \sqrt{{\tanh \left(\epsilon({\bf p},T_c)/
2T_c\right) \over \epsilon({\bf  p},T_c) }},
$$
$\zeta({\bf p})=\Delta ({\bf p})X({\bf p},T_c)$, and
$ H({\bf p},{\bf p}',T_c)= X({\bf p},T_c) X({\bf p}',T_c)
{\cal V}({\bf p},{\bf p}')$,
Eq.~(\ref{bcst}) is conveniently rewritten in the form of a linear
integral equation with a symmetric kernel,
\beq
\zeta({\bf p})=-{1\over 2}\int H({\bf p},{\bf p}',T_c)\zeta ({\bf p}')
d\upsilon' .
\label{eqh}
\eeq
Employing Eq.~(\ref{noze})), we may obtain the result
\beq
X({\bf p},T_c)=T_c^{-1/2}\sqrt{\dfrac{1-2n_*({\bf p})}{\ln\left((1-n_*({\bf p}))/n_*({\bf p})\right)}} , {\bf p}\in \Omega
\label{x}
\eeq
and observe that the function $X({\bf p},T_c)$ is greatly enhanced in
FC domains.

In view of the inverse proportionality to $T_c$ exhibited
by the kernel $H$, we infer that the overwhelming
contributions to the right side of Eq.~(\ref{eqh}) come
from the FC domains, provided the ratio $\eta/T_c$ exceeds
unity.  In that case, solution of this equation is
obviated, accounting for the minor change of the block
${\cal V}$ in a FC region.  Upon neglecting other
contributions to Eq.~(\ref{eqh}) and introducing a reduced
kernel $h({\bf p},{\bf p}')=(T_c/2) H({\bf p},{\bf p}')$,
we are left with the simple equation \beq T_c \zeta({\bf
p})=-\int_\Omega h({\bf p},{\bf p}')\zeta ({\bf
p}')d\upsilon', , \label{reqh} \eeq in which the kernel $h$
is practically $T$-independent.  We thus arrive at the
pivotal conclusion that $T_c$ changes {\it linearly} with
the strength of the interaction ${\cal V}$, a distinctive
fingerprint of high-$T_c$ superconductivity.  Further,
since the right side of Eq.~(\ref{reqh}) is, in fact,
proportional to the FC volume, we affirm that $T_c$ does
vary linearly with the FC density $\eta$, in agreement with
Eq.~(\ref{prop}).

In what follows we address the case of a 2D quadratic Brillouin zone
where a single small FC pocket resides in each quadrant of the zone.
Eq.~(\ref{reqh}) then reduces to the set of algebraic equations
\beq
T_c\,\zeta_l =-\sum h_{lk}\zeta_k  ,
\label{setb}
\eeq
where we have introduced quantities $\zeta_l=\zeta({\bf p}_l)$ and
$h_{lk}= h({\bf p}_l,{\bf p}_k)$, with indexes $l,k$ running from 1 to 4.
Obviously, $h_{ll}\equiv h_0$, $h_{l,l+1}\equiv h_1$, and
$h_{l,l+2}\equiv h_2$ are $l$-independent. Their signs and magnitudes
depend largely on the interplay between phonon attraction and Coulomb repulsion.
To solve the system (\ref{setb}) it is advantageous to make the substitution
$\zeta_k=e^{i\alpha k}$, with the requirement $e^{4i\alpha}=1$.  Analysis
demonstrates that there are 4 different high-$T_c$ solutions of the problem.
The first solution, corresponding to $\alpha=0$, where all $\zeta_k$ are
the same, occurs provided $ h_0+2h_1+h_2<0$.  It describes $S$-pairing,
with respective critical temperature $T_c=-( h_0+2h_1+h_2)$.

Another interesting solution, with
\beq
\alpha=\pi,\quad  T_c=2h_1-h_0-h_2,
\label{dp}
\eeq
has the usual $D$-pairing form, with $\zeta_1=\zeta_3=-\zeta_2=-\zeta_4$. It
occurs in the region of the Lifshitz phase diagram where $2h_1-h_0-h_2>0$.
Near the line of critical antiferromagnetic fluctuations where conventionally
$h_0=h_2=0$ while $h_1>0$, the solution (\ref{dp}) coincides with the
standard one. However, as seen, the presence of critical fluctuations
{\it is not a necessary condition} for the occurrence of high-$T_c$
$D$-pairing. In electron systems of solids hosting flat bands, this
solution can exist far from the critical line $T_N(x)$, even if the
$e-e$ interaction in the Cooper channel is repulsive.

A remaining pair of solutions, corresponding to $\alpha=\pi/2$ and
$\alpha=3\pi/2$ and having $\zeta_4=i\zeta_3=-\zeta_2=-i\zeta_1$,
describes $P$-pairing \cite{yak}, with the same critical temperatures
given by $T_c=h_2-h_0$.

In summary, we have generalized the Landau quasiparticle approach to
determine observable properties of systems possessing flat bands.
Specifically, we have motivated and introduced a set of two Landau-like
equations that replace the basic equation of Fermi liquid theory
connecting the single-particle spectrum and quasiparticle momentum
distribution.  Analyzing a typical phase diagram of the LCCO family of
electron-doped high-$T_c$ compounds, we have demonstrated that
gross properties of strange metals are incisively and economically
interpreted within the flat-band scenario developed here. Importantly,
we have shown that in a quadratic Brillouin zone, the BCS gap
equation has nontrivial solutions even if the interaction between
quasiparticles in the Cooper channel is of repulsive character.
The successful description of key features of the phase diagrams and
other properties of high-$T_c$ materials should provide ample
incentive for a change of theoretical course in the search for
deeper understanding of non-Fermi-liquid phenomena as well as
the mechanism of high-$T_c$ superconductivity.

The authors are indebted to V. Dolgopolov, Yu. Kagan, M.
Norman, J. Paglione, E. Saperstein, and J. Thompson for
valuable discussions.  This work is partially supported by
NS-932.2014.2 and by RFBR grants: 13-02-00085, 14-02-00044,
and 15-02-06261.  VAK thanks the McDonnell Center for the
Space Sciences for partial support.  JWC expresses his
gratitude to Professor Jos\'e Lu\'is da Silva and his
colleagues at Centro de Ci\^encias Matem\'aticas for
generous hospitality and fruitful discussions. VRS is
supported by the Russian Science Foundation, Grant No.
14-22-00281.

\end{document}